\documentclass[journal]{IEEEtran}
\def\svp{\Psi(v,\phi)}
\def\f{\frac}

\usepackage{graphicx}
\usepackage{color}

\ifCLASSINFOpdf
\else
\fi

\usepackage{hyperref}

\begin{document}
%\title{Probing the Birth of Space-Time Using Supercomputers}
\title{Glimpses of Space-Time Beyond the Singularities Using Supercomputers}

\author{\IEEEauthorblockN{
Parampreet~Singh$^\star$}\\
\IEEEauthorblockA{{\emph{Department of Physics \& Astrophysics, and,}\\ \emph{Center for Computation and Technology,} \\ \emph{Louisiana State University,} \emph{Baton Rouge, LA 70803, USA}}}\\
        % <-this % stops a space
\thanks{${}^\star$email: psingh@lsu.edu. This invited semi-technical overview article appeared in IEEE publication Computing in Science and Engineering, special issue ``Supercomputing-Enabled Advances in Science and Engineering" edited by S. Gottlieb and G. Khanna. (Jul/Aug issue 2018). 
 DOI Bookmark: \url{http://doi.ieeecomputersociety.org/10.1109/MCSE.2018.042781324}. Erratum-ibid: (Sep/Oct issue 2018). DOI Bookmark: \url{http://doi.ieeecomputersociety.org/10.1109/MCSE.2018.05329809}} % <-this % stops a space
}

\maketitle

\begin{abstract}
A fundamental problem of Einstein's theory of classical general relativity is the existence of singularities such as the big bang. All known laws of physics end at these boundaries of classical space-time. Thanks to recent developments in quantum gravity, supercomputers are now playing an important role in understanding the resolution of big bang and black hole singularities. Using supercomputers, explorations of the very genesis of space and time from quantum geometry are revealing a novel picture of what lies beyond classical singularities and the new physics of the birth of our universe.
\end{abstract}

%\begin{IEEEkeywords}
%IEEE, IEEEtran, journal, \LaTeX, paper, template.
%\end{IEEEkeywords}

% \fi
\IEEEpeerreviewmaketitle

%\section{Introduction}
%\IEEEPARstart{T}{his} demo file is intended to serve as a ``starter file''
%for IEEE journal papers produced under \LaTeX\ using
%IEEEtran.cls version 1.8b and later.
%I wish you the best of success.

%\IEEEPARstart{E}

\noindent
{\bf Introduction:} Einstein's theory of general relativity (GR) unifies the notion of classical space-time and gravity.
Its main lesson is that gravity is the dynamics of geometry of 4D space-time, and gravitational attraction occurs because the fabric of space-time gets curved due to the mass of objects. GR has been profoundly successful in describing the gravitational dynamics of bodies in our universe, and of the Universe itself. A recent notable example of this success was the observation of two binary black hole mergers, as detected by the Laser Interferometer Gravitational-Wave Observatory, which confirmed predictions of GR to a great accuracy. Despite remarkable success, it is widely believed that neither GR nor the notion of classical space-time are fundamental descriptions of nature because of the problem of singularities in the classical description of gravity.

Singularities are events during which gravitational attraction or the curvature of the space-time geometry diverges. An example is the big bang singularity, which occurs when our universe has vanishing volume. As a result, energy density of matter and space-time curvature explode to infinity. Such singularities are common in classical physics. In classical electrodynamics, Coulomb's law predicts that electric field due to a point charge is infinite at the location of the point charge. In the early days of GR, singularities were thought to arise as a result of certain assumptions in the model, similar to considering a point charge a mathematical abstraction. Big bang singularity was first found in the Friedmann-Lema\^{i}tre-Robertson-Walker (FLRW) cosmological model, which is homogeneous and isotropic, meaning that the geometry of the space is the same everywhere and in each direction, a reasonable approximation of our universe at large scales. Researchers initially believed that the big bang singularity in the FLRW model arose because of the drastic assumption of homogeneity and isotropy at all scales. But Penrose's and Hawking's powerful theorems, formulated in the 1960s, showed that singularities in GR do not occur under special circumstances, but rather they are generic features of Einstein's theory. At singularities, GR breaks down. Its failure to resolve singularities leads to an important question: Is there a more complete fundamental description of space-time in which singularities do not occur? As discussed below, the answer appears to be yes.

It has long been believed that singularities such as the big bang are a result of assuming classical continuum of space-time at all scales, and will be resolved when space-time is quantized in a quantum theory of gravity -- a marriage of classical gravity and quantum theory. A fundamental lesson from the latter is that physical quantities that classically take continuous values will, upon quantization, take discrete values. Classical physics is an approximation of the limit at which quantum discreteness vanishes. In a quantum world, an electric charge can't be localized to a point, and naturally, there is no classical singularity of the electric field. A fundamental question is, can quantum discreteness similarly resolve space-time singularities? For big bang and black hole singularities, the volume of the spatial region vanishes, causing space-time curvature to blow up. If in quantum gravity, space-time geometry is not continuous but rather discrete with a nonvanishing minimum volume, then the problem of space-time singularities can be successfully addressed.

As we will see, departure from the classical continuum space-time to quantum discrete space-time brings many challenges to the extractions of physics using numerical simulations. Whereas various numerical simulations to address interesting problems in GR could be performed on a single core, and HPC is used to tackle complex problems, the situation in quantum gravity is quite different. To answer even the simplest questions, using supercomputers becomes necessary. In recent years, tools have been developed to overcome the challenges associated with performing simulations on quantum discrete space-times, and many numerical simulations using HPC have been performed. The resulting physics is strikingly different from GR in the sense that there is no big bang singularity when the quantum discreteness of space-time is considered.\\

\noindent
{\bf From classical to quantum geometry:} One of the main candidate theories of quantum gravity is loop quantum gravity (LQG) \cite{thiemann,rovelli,aa-jp}. Unlike other approaches to quantum gravity, it is nonperturbative and background independent. In simple words, it means that LQG treats gravity as dynamics of space-time in the true spirit of Einsteinian gravity, not just as another force on a spectator space-time, which is a central theme of Newtonian gravity and other fundamental forces. Conventionally, quantization of GR, such as in the Wheeler-DeWitt theory -- named after pioneers John Wheeler, who made many seminal contributions to GR and quantum gravity and also coined the term black hole, and Bryce DeWitt, a founding father of canonical quantum gravity approach -- has been studied using the metric and its momentum as basic variables. A metric quantifies distance between objects on spatial geometry, and its momentum tells us the way the metric changes under time evolution. It turns out that the resulting Hamiltonian, a primary quantity that reveals a system’s dynamics and energy levels, in these variables is unmanageable at the quantum level. In 1980's, Ashtekar found that instead of the metric, if one considers triads and their momentum (connection), then the Hamiltonian is manageable \cite{abhay}. Triads are just a different way to capture spatial geometry encoded in the metric through three orthogonal vector fields; connection, which is conceptually similar to the vector potential in electrodynamics, captures the way geometry changes over time. A decade of rigorous mathematical work in the 1990's showed that the resulting theory, LQG, is kinematically different at the quantum level from the Wheeler-DeWitt theory. Instead of the classical continuum space-time of GR and the Wheeler-DeWitt theory, the quantum space-time turns out to be discrete in LQG. The classical differential geometry is replaced by a quantum geometry in which geometrical operators such as area and volume have discrete spectrum with nonvanishing minimum eigenvalues.

A consequence of quantum geometry is the boundedness of energy density and space-time curvature. This is straightforward to understand by recalling that in quantum mechanics, if one of the phase space variables is discrete, its conjugate variable is bounded. It turns out that energy density and space-time curvature are conjugate to geometric operators that have a discrete spectrum in LQG. Since the space-time curvature is bounded, in cosmological space-times that have been rigorously quantized using LQG, quantum discreteness results in a nonsingular evolution. An important caveat is that all the loop quantized space-times so far are homogeneous. Nevertheless, they are still quite nontrivial, including the FLRW space-time capturing the dynamics of our universe.

Important features of the quantum evolution of the above space-times in LQG include the following. Unlike the differential equations of GR, the evolution equation in LQG is a finite difference equation with discreteness fixed by the underlying quantum geometry. In particular, the Hamiltonian is a finite difference equation with a uniform spacing in volume of the Universe. As is expected from a consistent quantum gravitational theory, it approximates the Hamiltonian in GR, which is a differential equation, when space-time curvature is much smaller than the Planck curvature (defined as $c^3/\hbar G \approx 3.83 \times 10^{65} \mathrm{cm}^{-2}$). When such a high curvature is reached, there are important differences between the two. As a result, if one considers a quantum state peaked on a classical expanding solution of GR and evolves it backward toward the big bang using LQG, the state follows the classical trajectory for a long time but shows significant departures when space-time curvature becomes very large. Near the Planck curvature, the volume of the universe stops shrinking, and starts increasing. As a result, the big bang is replaced by a turnaround of volume -- a nonsingular big bounce! The bounce is caused by quantum gravitational repulsion at Planck scale resulting from the discreteness of quantum geometry. Unlike in GR, energy density and space-time curvature remain finite in LQG. Interestingly, there exists an effective continuum description with modifications to differential equations of GR that capture the loop quantum dynamics quite successfully, at least for quantum states that bounce at volumes much larger than the Planck volume. These novel results were first obtained for isotropic and homogeneous space-times about a decade ago using numerical simulations performed on a single core \cite{aps1,aps3}. Avoidance of the big bang and occurrence of the big bounce have since been shown to be a common feature of various cosmological models based on LQG \cite{as-status}, using various analytical and numerical techniques \cite{numlqc1,numlqc2}.

If the big bounce truly reflects the fundamental physics of the very early universe, then this prediction of LQG must pass some stringent robustness tests. First, how generic is the bounce in different cosmological and black hole space-times. In particular, does the bounce occur for anisotropic space-times that capture the generic approach to singularities? Second, does the bounce occur for generic states or is it a feature of only specific quantum states? Finally, the bounce can potentially leave invaluable signatures of LQG in cosmic microwave background (CMB) and primordial gravitational wave background originating in the very early universe. For this, it is important to understand the regime of validity of effective continuum space-time description. To answer the above questions, and to potentially connect LQG with CMB and primordial gravitational wave observations, certain computational challenges associated with performing numerical simulations on quantum geometry must be overcome.\\

\noindent
{\bf Computational challenges of quantum geometry:} The quantum Hamiltonian in LQG is a difference equation in the variables capturing spatial geometry, an example being the spatial volume of the universe in the FLRW model. The discreteness in the difference equation is completely determined by LQG, with no parameter freedom to change it. In contrast, in GR, the fundamental equations are differential and finite difference equations are used as approximations in numerical computation with a freedom to vary discreteness for better accuracy. But, in LQG numerical simulations must carefully take in to account the underlying quantum discreteness which fixes the allowed numerical grid. In particular, for a stable evolution, the Courant-Friedrichs-Lewy (CFL) condition must be satisfied \cite{cfl}. In the continuum limit, where the quantum discreteness becomes negligible, the difference equation results in a hyperbolic partial differential equation, which is the Wheeler-DeWitt equation in Wheeler-DeWitt quantization of cosmological space-times. In the latter, the space-time is continuous as in GR, and states evolved using Wheeler-DeWitt equation follow classical trajectory of GR at all times \cite{aps3}. The CFL condition implies that given a discreteness in spatial grid, the temporal grid discreteness must be small enough such that the numerical speed of propagation is greater than the characteristic speed in the Wheeler-DeWitt equation. For a stable evolution, the fixed discreteness poses certain challenges and results in the demand for huge computational resources. 

We illustrate this for the case of isotropic cosmological model sourced with a massless scalar field $\phi$ - a toy model to describe matter, whose strength varies monotonically. The latter property allows $\phi$  as a clock to study quantum dynamics of the universe. In canonical description of gravity, physical solutions are obtained from a constraint which the Hamiltonian must satisfy. The ``energy balance'' of matter and gravity requires that it must vanish. This Hamiltonian constraint using LQG turns out to be a difference equation in volume of the spatial slices of the universe denoted by $v$ \cite{aps3},
\begin{eqnarray}\label{iso-eq}
\partial_\phi^2 \svp &=& B^{-1}(v)(C_+(v) \Psi(v + 4, \phi) + C_0(v) \svp \nonumber \\ && ~~~~~~~~~~~~~~~~~~~  + C_-(v) \Psi(v - 4)),
\end{eqnarray}
where $C_+$, $C_0$, $C_-$  and $B$ are determined by the action of geometrical operators in quantum theory, with eigenvalues:
\begin{equation}
 C_-(v) = \nonumber C_+(v-4) = \f{\pi G}{4 \times 3^{3/4}} \, |v-2| \left||v-3| - |v-1|\right|, %~~ C_-(\nu) = \f{3 \pi G}{4 \ld^2} \, \nu (\nu - 2 \ld), ~~ C_0(\nu) = - \f{3 \pi G}{ \ld^2} \, \nu ~
 \end{equation}
\begin{equation}
C_0(v) = \nonumber -C_+(v) - C_-(v)
\end{equation}
and
\begin{equation}
 B(v) = \nonumber \f{3^{5/4}}{4} |v| ||v+1|^{1/3} - |v-1|^{1/3}|^3 ~.
\end{equation}
This equation couples the wavefunction $\Psi$  of the universe in uniform steps of four times the Planck volume. In quantum theory, $\Psi$   plays a central role in providing information about the values a physical quantity takes through computation of expectation values. Using $\Psi$, expectation values of volume operator at different times can be computed, finding exactly when the universe bounces. Eq. \ref{iso-eq} results in a stable evolution and classical solutions at late times \cite{numlqc1}. At large volumes it yields the Wheeler-DeWitt equation
\begin{equation}
 \frac{\partial^2\Psi}{\partial\phi^2} = 12\pi G v\left (
  \frac{\partial}{\partial v}\left ( v\frac{\partial\Psi}{\partial v}\right )
  \right ) %= 12 \pi G \frac{\partial^2\Psi}{\partial x^2}.
\end{equation}
which has characteristic speeds: $\lambda^{\pm} = \pm \sqrt{12\pi G} v$. The CFL condition constrains the maximum time step $\Delta \phi$ as $\Delta \phi \le 4/|\lambda^{\pm} | \propto v^{-1}$.  Thus, a large spatial grid requires a very fine time grid. Therefore, investigating cosmological space-times in LQG turns out to be very expensive.
As an example, on a single core a typical simulation with spatial grid of $10^{12}$ volume in Planck units requires approximately $10^{10}$ years \cite{chimera}.

\begin{figure*}[tbh!]
\centering
{\includegraphics[width=3.5in]{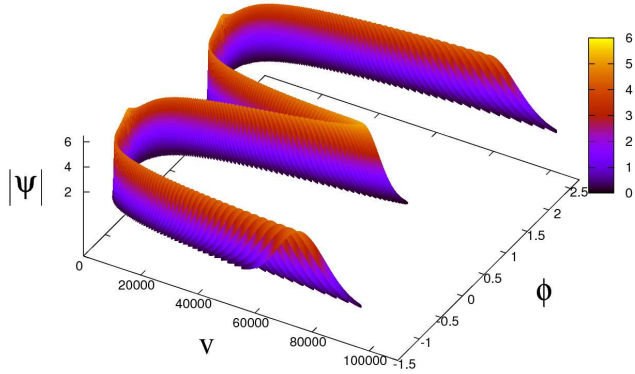}%
\label{fig1}}
%\hfil
%\subfloat[Case II]{\includegraphics[width=2.5in]{box}%
%\label{fig_second_case}}
\caption{From big bang to big bounce. The magnitude of the wavefunction of the universe is plotted versus the volume of the isotropic  universe, while the scalar field $\phi$ plays the role of time. The big bang and big crunch are avoided in the quantum theory as the plot demonstrates.   This figure corresponds to  the case of a cyclic universe which classically encounters the big bang singularity in the past and the big crunch singularity in the future.   Only a snapshot of the evolution of the state in a region near the classical singularity is shown.   }
%\label{fig_sim}
\end{figure*}

This computational cost dramatically increases for anisotropic cosmological space-times such as Bianchi models, for which geometry, in contrast to FLRW space-time, in each direction is different,  and for black holes which have a central singularity similar in properties as the big bang singularity.  Bianchi spacetimes are important to study as they are believed to capture the generic approach to singularities in GR \cite{bkl}.
If matter is absent then the numerical grid is 3-dimensional, composed of three directional volumes $v_i$, measuring spatial volume in each direction. Of these one (or its conjugate variable) can be used as a `clock' to measure the other two spatial directions. Unlike the example of the isotropic cosmology where temporal grid is determined by the CFL condition, here numerical grid in all the directions is completely fixed by quantum geometry. The number of spatial grid points for the computations grows by $N^2$, where $N$ is the number of grid points
required in the isotropic case. As discussed later, a typical simulation for quantum states which are sharply peaked require about $10^{15}$ floating point operations and about 500 GB of memory. For wider states, and states probing deep quantum geometry, typical simulations require $10^{19}$ floating point operations with memory requirements exceeding $\sim$ 5 TB. For Bianchi space-times with a scalar field, one deals with a three dimensional spatial grid and a temporal grid in the scalar field whose discreteness is constrained by the CFL condition. In this case, the number of floating point operations climb to $10^{25}$ for a typical simulation involving just sharply peaked states.

To answer questions about the resolution of singularities and probing the deep Planck regime to understand the emergence of the space-time as we know it from quantum geometry, one needs to perform many simulations such as above with a wide
range of parameters. Given the
 high computational cost of performing numerical simulations on quantum geometry, especially for anisotropic space-times, we must bring forth smarter algorithms which can be efficiently used in high performance computing.  In the following we first describe one such algorithm, the Chimera method \cite{chimera}, which has been successfully implemented to understand singularity resolution for a wide variety of states including those which have very wide spreads, squeezing and non-Gaussian properties \cite{numlsu-2,numlsu-3}. Next, we describe the computational implementation for investigating the resolution of singularities in anisotropic and black hole space-times, where above algorithms can be used. \\

\noindent
{\bf Chimera -- a hybrid of quantum and classical geometry grids:} The Chimera algorithm \cite{chimera}  reduces the computational cost for numerical simulations on quantum geometry by using some of the key properties of the quantum Hamiltonian obtained for isotropic and anisotropic space-times in LQG.
 As noted earlier, at large volumes compared to the Planck volume, which also correspond to small space-time curvatures, the quantum Hamiltonian is approximated extremely well by the Wheeler-DeWitt equation. Since Wheeler-DeWitt states are peaked on classical solutions of GR, one thus finds that classical continuous space-time emerges from the quantum geometry when the space-time curvature becomes much smaller compared to the Planck curvature scale. In fact, the eigenfunctions of the quantum Hamiltonian obtained using LQG are found to be superposition of the eigenfunctions of the Wheeler-DeWitt equation at large volumes \cite{aps3}. Further, the most non-trivial quantum gravitational effects are concentrated in the regime close to the bounce. At the bounce, the eigenfunctions decay exponentially and vanish near the singularity \cite{aps3,craig}. Thus, if one can consistently split the numerical grid in two parts: an inner grid where finite difference equation from LQG is used, and an outer grid where one can approximate the finite difference equation using Wheeler-Dewitt equation, then above observations suggest negligible difference in numerical simulations carried over a full quantum geometric grid and the one where only inner grid is quantum geometric. The interface of the inner and outer grids is to be located at a carefully chosen large volume. It should be large enough such that the approximation is excellent and does not introduce any numerically significant errors. On the other hand the interface should be at a small volume such that the constraints from the CFL condition are alleviated. An additional input further reduces the computational cost by changing the time discretization through the CFL condition. Since the outer grid is not constrained by the quantum geometry, one is free to do a coordinate transformation $x = \ln v$ as a result of which the characteristic speeds for the corresponding evolution equation in $x$ become $\lambda^\pm = \pm \sqrt{12 \pi G}$. The CFL condition then implies that the maximum time step $\Delta \phi$ is proportional to the discreteness in volume only and is independent of the size of the outer grid. This change in coordinates brings
 a significant reduction in the computational cost because of the following reasons. On the outer grid, since the coordinate is logarithmic one needs less refined spatial grid, and even if the outer grid is very large the discreteness in temporal grid does not need to be made smaller for stability. Further reduction in computational resources can be achieved by using a Discontinuous Galerkin method to approximate derivatives on the outer grid. Using such a higher order scheme one can perform numerical
 simulations with same accuracy as before with a much lower resolution. Rigorous analysis with different interface boundaries and a variety of quantum states shows that the usage of two different grids to solve finite difference quantum Hamiltonian is quite successful in significantly reducing the computational cost \cite{chimera,numlsu-2,numlsu-3}.

 An example of the application of the Chimera method described above in conjunction with HPC is shown in Fig. 1, which shows a bounce for a quantum state initially peaked at large volume instead of encountering the big bang singularity at vanishing volume. The role of time is played by the scalar field $\phi$.  The model corresponds to a cyclic universe with a negative potential term in eq. (\ref{iso-eq}) which causes a classical turn-around at large volumes. In GR, cycles are not possible because of  big bang singularity.  Starting from positive values of $\phi$, in the backward evolution the state turns around at very small volume due to quantum gravity effects, avoiding the big bang and undergoes a bounce to an expanding branch at $\phi \sim 2$ (in Planck units). After the classical turnaround at large volumes occurring at $\phi \sim 1$, the state evolves again towards a big bang which is once again avoided by quantum geometry. The cycle repeats in further evolution. \\

\noindent
{\bf Supercomputing implementation for quantum space-times:} In the above discussion we saw the way Chimera method \cite{chimera} can cut the computational cost significantly. The method can be used both for isotropic and anisotropic space-times. Recall that for latter, spatial geomtery in different directions is quite different which has many consequences. First, anisotropies dictate the structure of a generic singularity. The big bang in an anisotropic universe may not be a point as in FLRW model, but a cigar because of the way different directions contract. Further, for our universe to isotropize from anistropic initial conditions is  non-trivial, and there can be phenomenological signatures of this process occurring in very early universe in CMB and primordial gravitational waves.  On the numerical side, computational cost of studying such space-times is high because of the increase in number of distinct directions to three captured by three directional volumes $v_i$.  Here Chimera method plays a supplemental role in the main computational kernel. We now describe the primary elements of the computational algorithm  (see \cite{numlsu-4} for details).

As for the case of the FLRW model,  loop quantization of the Bianchi-I space-time yields the quantum Hamiltonian constraint as the following difference equation \cite{madrid}:
\begin{equation}\label{C-q}
\hat{\mathcal{H}}= -\frac{2}{\gamma^2}\left( \hat{\Theta}_1\hat{\Theta}_2 +
\hat{\Theta}_1\hat{\Theta}_3+\hat{\Theta}_2\hat{\Theta}_3\right) ~ \approx 0,
\end{equation}
where $\approx 0$ indicates that the physical solutions are obtained from the vanishing of the Hamiltonian constraint. Here $\hat{\Theta}_i$ have the following action of the eigenstates corresponding to directional volumes $v_i$
\begin{equation}
\hat{\Theta}_i|v_i\rangle=-i \frac{\Delta_{\mathrm{Q}}}{2\sqrt{3}}\left( f_+(v_i)|v_i+2\rangle-f_-(v_i)|v_i-2\rangle \right),
\end{equation}
where $\Delta_{\mathrm{Q}} \approx 1.35 \times 10^{-65} \mathrm{cm}^2$ is the minimum eigenvalue of the area operator computed in LQG, and
\begin{equation}
f_\pm(v_i)=g(v_i\pm 2) s_{\pm}(v_i) g(v_i)
\end{equation}
with
\begin{equation}
s_{\pm}(v_i)= \mathrm{sgn}(v_i\pm 2) + \mathrm{sgn}(v_i)
\end{equation}
and
\begin{equation}
g(v_i)= \left| \left|1+\frac{1}{v_i}\right|^{1/3}
             \left|1-\frac{1}{v_i}\right|^{1/3}\right|^{-1/2}
\end{equation}
if $ v_i\neq 0$, and is zero otherwise. The uniform discreteness is of two Planck units of directional volumes $v_i$. Physical states
can be constructed using the eigenfunctions $e_{{\omega_i}}$ of $\hat \Theta_i$ operators with eigenvalues $\omega_i$ which capture the way anisotropy changes in different directions. Using the operator $\hat \Theta_i$, one finds that eigenfunctions satisfy following relations:
\begin{equation}
e^{\epsilon_i}_{\omega_i}(2+\epsilon_i) =
-i\frac{\sqrt{3}\omega_i}{\Delta_{\mathrm{Q}}} \frac{e^{\epsilon_i}_{\omega_i}(\epsilon_i)}{g(2+\epsilon_i)g(\epsilon_i)},
\end{equation}
and
\begin{eqnarray}
e^{\epsilon_i}_{\omega_i}(2n+2+\epsilon_i) &=& \nonumber
\frac{g(2n-2+\epsilon_i)}{g(2n+2+\epsilon_i)}
e^{\epsilon_i}_{\omega_i}(2n-2+\epsilon_i) \\ && -i\frac{\sqrt{3}\omega_i}{\Delta_{\mathrm{Q}}}
\frac{e^{\epsilon_i}_{\omega_i}(2n+\epsilon_i)}{g(2n+2+\epsilon_i)g(2n+\epsilon_i)} \nonumber \\
\end{eqnarray}
where $n > 0$ and labels the lattice of $v_i$, and $0 < \epsilon_i \leq 2$. Using above recursion relations we can evaluate the wavefunction in the entire range of $v_i$'s starting from some initial values.

To extract physical predictions, one can study the relational dynamics of directional volumes $v_2$ and $v_3$ with respect to $v_1$ or its conjugate $b_1$.  The latter turns out to be the preferred choice, because unlike $v_1$ it is monotonic which is a required property of a good clock. The physical state for any fixed lattice $\epsilon_i$, with $b_1$ playing the role of the clock, is
\begin{equation}\label{phys-state}
{ {\Psi}_{b_1}}(v_2,v_3)=\int {\rm d}\omega_2 {\rm d}\omega_3 \tilde \Phi{(\omega_2,\omega_3)} e_{\omega_1}(b_1) e_{\omega_2}(v_2) e_{\omega_3}(v_3),
\end{equation}
where $\tilde \Phi$ provides the profile of the quantum state, chosen to be a Gaussian peaked at $\omega_2^*$ and $\omega_3^*$, with spreads $\sigma_2$ and $\sigma_3$:
\begin{equation}
\tilde{
  \Phi}(\omega_2,\omega_3)=
%\frac{e^{-(\omega_2-\omega_2^{\star})^2/2\sigma_2^2}}{\sqrt{\pi}\sigma_2}
%e^{i\beta_2\omega_2}
\frac{1}{\sqrt{\pi}\sigma_2}
e^{-\frac{(\omega_2-\omega_2^*)^2}{2\sigma_2^2}} e^{i\beta_2\omega_2} \,
\frac{1}{\sqrt{\pi}\sigma_3}
e^{-\frac{(\omega_3-\omega_3^*)^2}{2\sigma_3^2}} e^{i\beta_3\omega_3} ~.
% should it be \sqrt{2\pi} instead?
\end{equation}
We can then obtain the expectation values of $\hat v_2$ and $\hat v_3$ which act as multiplication operators.

\begin{figure*}[tbh!]
\centering
{\includegraphics[width=3in]{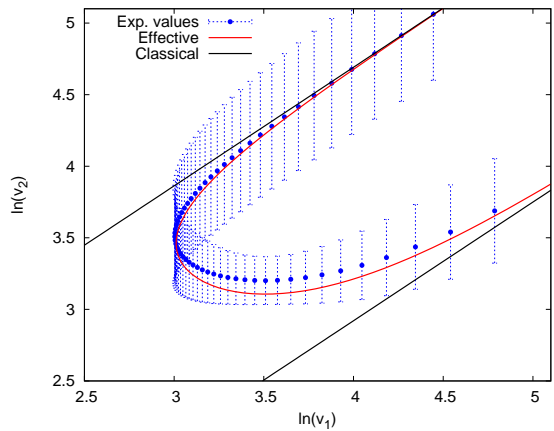}%
\label{fig2}}
%\hfil
%\subfloat[Case II]{\includegraphics[width=2.5in]{box}%
%\label{fig_second_case}}
\caption{Bounce in anisotropic space-time. This plot shows the bounce of a quantum state initially peaked at the classical trajectory (upper solid black curve) in logarithmic variables in two directions of the Bianchi-I space-time. As before only the region close to singularity is shown. The classical curves are singular and are disconnected. Quantum gravity effects cause a bounce of the state from one curve to the other. The quantum expectation values, shown by black dots, and corresponding dispersions are captured extremely well by an effective space-time trajectory (red curve).  }
%\label{fig_sim}
\end{figure*}

With $b_1$ playing the role of clock, for each value of $b_1$, the physical state which is a three dimensional object, can be stored as an array of size $n_2 \times n_3$, where $n_2$ and $n_3$
are the sizes of the spatial grid in $v_2$ and $v_3$. At any given time step, computation of a physical state can be parallelized in $v_2$ and $v_3$ directions. If the grid in $b_1$ is labeled by $n_1$, then to compute the physical state and expectation values over this time interval, we need to evaluate integrals  in eq.(\ref{phys-state}) $n_1 \times n_2 \times n_3$ times. $\tilde \Phi(\omega_2,\omega_3) e_{\omega_1}$ is a double precision complex array of size  $n_{\omega_2} \times n_{\omega_3} \times n_1$ which is computed initially using discrete FFT and is stored as a vector of grid arrays distributed among processors in $\omega_2$ and $\omega_3$ directions but not in $b_1$ direction to compute the physical state and expectation values. This particular array requires a large amount of storage. As an example, for $n_1 = 131072$ and $n_{\omega_2} = n_{\omega_3} = 256$, one requires storage of 128 GB. Using MPI, computation of FFTs is parallelized by evaluating FFTs serially for different values of $\omega_2$ and $\omega_3$ on different processors.  Integrals are computed using Gauss-Legendre quadrature with eigenfrequencies in range $\omega_i^* \pm 10 \sigma_i$. At each time step, we have a double loop over spatial grid in $v_2$ and $v_3$, and for each spatial grid point  there is a double loop in $\omega_2$ and $\omega_3$. Including all steps, 22 multiplications and 22 additions are required in each loop. The number of floating point operations are thus:
 \begin{equation}\label{Neq}
 N = 44 \times n_1 \times n_2 \times n_3 \times n_{\omega_2} \times n_{\omega_3} ~.
\end{equation}
%\be
%{ {\tilde \chi}}_{b_1}(\omega_2,\omega_3)=\tilde{  \Phi}(\omega_2,\omega_3) {\tilde e}^\prime_{\omega_1}(b_1)
%
MPI parallelization is done using Cactus framework \cite{cactus,cactus_web}, which is extensively used in numerical relativity. In Cactus, the central core called ``flesh'' is connected to different modules called ``thorns.'' Cactus implementation for solving finite
difference quantum Hamiltonian in LQG was accomplished by introduction of a new ``thorn'' for evaluating physical state and expectation values of relevant operators \cite{numlsu-4}. The LQG thorn works in conjunction with other computational thorns and the flesh. Outer loops are parallelized with OpenMP and the inner loops are written such that they can be auto vectorized by compilers. The main computational kernel has been ported to GPUs using OpenACC, and to Intel Xeon Phis using OpenMP with Intel's offload compiler directives.  In both cases the code can use both CPUs and accelerators at the same time by dynamical adjustment of the computational work.

The computational kernel achieves $ 60 \%$ of the theoretical peak performance on the single core. On a 16 core shared memory node using OpenMP parallelization, kernel achieves about $50 \%$ of the peak performance using CPUs. On the other hand, using Intel's Xeon Phi accelerator cards the performance is about $20 \%$ due to data cache misses. The performance on GPUs is about $25-30\%$.  Strong and weak scaling tests on Stampede supercomputer at Texas Advanced Computing Center using XSEDE resources \cite{xsede} give excellent results \cite{numlsu-4}.  On Stampede, for strong scaling, increasing the nodes from $3$ to $256$ (with each node having 16 cores and 1 Xeon Phi) increases the speedup $68$ times, less than the ideal case
where the speedup should be $85.33$ times. Here the speedup is the ratio of the time taken to perform the same computation using 3 nodes to the time taken with 256 nodes.  For the weak scaling, as the number of nodes are increased the grid size $n_2 \times n_3$ is increased accordingly keeping other parameters fixed. Increasing the nodes from 3 to 1024, the code slows down by only $10 \%$.

In a typical simulation, a quantum state is considered at large volumes peaked at the classical trajectory. This state serves as an input to determine the physical state (\ref{phys-state}) allowed by the quantum Hamiltonian constraint (\ref{C-q}). Results from a typical simulation are shown in Fig. 2. The plot corresponds to a sharply peaked Gaussian state with $\omega_2^* = 100$, $\omega_3^* = 1000$, with spreads $\sigma_2 = 14$ and $\sigma_3 = 40$. For this state, one requires $n_1 = 2^{11}$, $n_2 = n_3 = 2^{12}$ and $n_{\omega_2} = n_{\omega_3} =  2^8$. Using eq.(\ref{Neq}), the total number of required floating point operations are  $N = 44 \times 2^{51} \approx 10^{17}$ flops. In the figure, classical trajectories shown by solid black curves are compared with the expectation values (shown with black dots) of the logarithm of directional volume $v_2$  plotted versus expectation values of logarithm of directional volume $v_1$, along with the dispersions. In the classical theory, there are two solutions for the above values of parameters which are both singular and disjoint. Starting from any of the classical curves, a big bang singularity is encountered at vanishing values of $v_1$ and $v_2$. The quantum state is chosen at the large values of directional volume, peaked at the upper classical curve. The quantum state follows the classical curve for a long time until it reaches the Planck regime where departures from the classical theory become significant. Instead of going towards the singularity, the quantum state bounces from the singular classical solution towards another classical solution. Thus the classical singularity is avoided in LQG. Interestingly, using coherent states the finite difference quantum Hamiltonian constraint (\ref{C-q}) can be approximated by
a differential effective Hamiltonian constraint which captures the quantum gravitational effects quite well. The resulting effective Hamiltonian is given by \cite{numlsu-4}
\begin{eqnarray}
{\cal H}_{\mathrm{eff}} \hskip-0.3cm &=& \hskip-0.3cm - 72 \pi \ell_{\mathrm{Planck}}^4 \bigg[\sin(b_1) v_1 \sin(b_2) v_2  \nonumber \\ && \hskip-0.5cm + \sin(b_2) v_2 \sin(b_3) v_3  + \sin(b_3) v_3 \sin(b_1) v_1 \bigg] .
\end{eqnarray}
Note that the above Hamiltonian is significantly different from the classical Hamiltonian which is obtained in the limit when $b_i$ the conjugate variables to $v_i$ are very small. It captures the discrete quantum gravitational effects
in an effective continuum space-time. In a sense, it can be used to extract quantum dynamics in terms of the variables of the classical continuum space-time. We see from Fig. 2, that the effective dynamical trajectory obtained from the above Hamiltonian matches quite well with the classical curves at late times, and with quantum expectation values at all the times within the value of dispersions. As with the quantum theory, the effective dynamics captures the singularity resolution.

\begin{figure*}[tbh!]
\centering
{\includegraphics[width=3in]{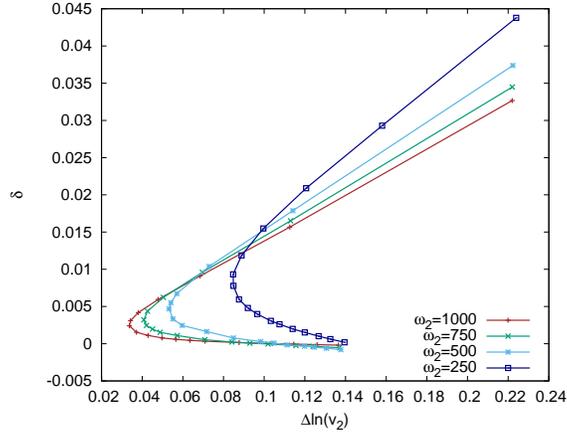}%
\label{fig3}}
%\hfil
%\subfloat[Case II]{\includegraphics[width=2.5in]{box}%
%\label{fig_second_case}}
\caption{Validity of the effective space-time description.  The relative difference in the expectation value of $\ln(v_2)$ at the bounce and the value predicted by the effective theory are plotted for various simulations. Each dot corresponds to a simulation performed using LQG thorn in Cactus.  For each value of  $\omega_2$ at which the state is peaked, a non-monotonic behavior is found. The difference between quantum theory and effective theory decreases as the spread in the state decreases only up to a certain value. The lowest values of $\delta$ are found for larger spreads. The non-monotonic behavior, cross-over between different curves and change in sign of $\delta$ reveals a non-trivial dependence on the value of $\omega_2$ and the spread. Each dot corresponds to one numerical simulation with at least $10^{16}$ flops. }

%\label{fig_sim}
\end{figure*}

Since many phenomenological predictions are extracted from effective dynamics, understanding the validity of effective space-time is an important problem in LQG. As we mentioned above, effective space-time provides a continuum description of underlying quantum geometry up to the scale of singularity resolution. If we can understand at what scales the effective dynamics is trustable, we can then understand the scale at which the continuum space-time emerges from quantum gravity and the phenomenological predictions are reliable. A measure of this validity is to explore the prediction about the volume at which the bounce occurs in the quantum dynamics and the effective space-time description. Due to the wide range of volume between the classical singularity and the bounce regime in LQG, for a better visualization it is helpful to plot the relative difference in logarithm of volumes. This relative difference in $v_2$ direction  is $\delta = (\ln(v_2) - \ln(v_2)^{\rm{eff}})/\ln(v_2)^{\rm{eff}},$ computed at the bounce of directional volume $v_2$. If this difference is small, the effective dynamics can be trusted and the resulting physics, which includes signatures in CMB are reliable. It was earlier thought that decreasing the spread in volume would cause a better agreement between the quantum theory and the effective dynamics because the state will be more sharply peaked in volume and any effects from quantum fluctuations will not cause significant errors between quantum and effective dynamics.   Thanks to the extensive
numerical simulations carried out for a large set of parameters, the actual picture turns out to be strikingly different. Fig. 3 shows that relative difference in the bounce volume $v_2$ in logarithmic variables for various values of eigenfrequency $\omega_2$ on which the quantum state is peaked plotted versus absolute fluctuation in the logarithm of the volume $v_2$. Each point in the plot corresponds to a numerical simulation involving more than $10^{16}$ flops. We find some surprising results. First, for each value of $\omega_2$ at which the state is peaked the magnitude of relative difference between the bounce volume between quantum and effective theory is not smallest for smallest $\Delta \ln (v_2)$, rather is is smallest for some larger value. That is, the agreement between quantum dynamics and effective description does not increase monotonically when the state becomes more peaked. This shows that the conventional wisdom of increasing agreement between quantum geometry and effective space-time description for states which are more sharply peaked needs revision. In fact, for any given value of $\omega_2$, there is a minimum allowed value of $\Delta \ln (v_2)$.
 The magnitude of relative difference between quantum and effective dynamics, $\delta$, shows a non-monotonic behavior for each value of $\omega_2$ at which the state is peaked. Each curve has turnaround point, which for larger values of $\omega_2$ occurs at smaller values of $\delta$. Interestingly, for large enough values of the spread, the relative difference becomes negative. All this shows that the reliability of the effective dynamics depends on the quantum state in a non-trivial way. For most cases, the effective dynamics causes a bounce at smaller volume than the quantum dynamics making $\delta$ positive. In all such cases, effective dynamics underestimates the quantum gravitational repulsion. However, there exist cases of large fluctuations where the opposite happens and effective dynamics overestimates quantum effects. These are important clues to the reliability of effective dynamics and the new physics at the Planck scale in LQG.  We  emphasize that all these results are possible only because of the extensive numerical simulations with HPC for loop quantized Bianchi-I space-times. The analytical understanding of turnaround, negative values of $\delta$ and non-monotonic behavior in Fig. 3  is yet to be fully established. In fact, the latter behavior was not anticipated through analytical studies so far. This is an example where numerical simulations lead to discovery of the new physics in the Planck regime where detailed analytical understanding is still lacking. \\

\noindent
{\bf Summary:} In GR, singularities are the final boundaries at which evolution stops and all known laws of physics break down. The hope has been that a quantum theory of gravity will eliminate these boundaries, extending the space-time beyond the big bang. But understanding properties of quantum space-time is a very hard problem. At the analytical level, the theory is yet to be fully deciphered, and direct observational tests are absent. However, thanks to the significant analytical understanding of quantization in LQG achieved in the past three decades, interesting cosmological and black hole space-times can be loop quantized and singularity resolution can be studied. In LQG, quantum geometry brings forth computational challenges never before encountered in GR. Supercomputers become necessary to answer even the most basic questions, and play an important role in deciphering the physics of quantum space-time. The key prediction for loop quantized isotropic cosmological and anisotropic space-times is the absence of a big bang, which is replaced by a big bounce. For loop quantum black hole space-times, similar results are emerging \cite{bh-quantum}. These results radically change our understanding of the origin of our universe and the central singularity of black holes. If the prediction of bounce holds, then LQG tells us that there existed a large universe before what we observe as the big bang. Numerical studies using HPC reveal the existence of an effective space-time description that sheds important light on the way continuum space-time emerges from quantum geometry and potentially links LQG with astronomical observations. In coming years, one challenge is to extend these results to inhomogeneous space-times where the understanding of analytical aspects in quantum gravity is yet to be completed. Given the progress over the past couple of years, it can be expected that supercomputers will prove to be an invaluable and essential tool for the complete discovery of the new physics at the Planck scale, and to go beyond the limitations of Einstein's GR.

\section*{Acknowledgment}
I am grateful to Peter Diener, Brajesh Gupt, Anton Joe and Miguel Megevand for collaboration. I thank anonymous referees for valuable comments on improving presentation of this manuscript.
This work is supported by  NSF  grant  PHY-1454832.  This work used the Extreme Science and Engineering Discovery Environment (XSEDE),
which is supported by National Science Foundation grant number ACI-1053575, and HPC resources at Center for Computation and Technology at Louisiana State University.

\ifCLASSOPTIONcaptionsoff
  \newpage
\fi


\begin{thebibliography}{1}

\bibitem{thiemann} T. Thiemann, \emph{Modern Canonical Quantum General Relativity}, Cambridge Monographs on Mathematical Physics,
Cambridge University Press, 2008.

\bibitem{rovelli} C. Rovelli, \emph{Quantum Gravity}, Cambridge University Press, 2007.

\bibitem{aa-jp} A. Ashtekar, J. Pullin, \emph{Loop Quantum Gravity: The First 30 Years (100 Years of General Relativity)}, World Scientific, 2017.

\bibitem{abhay} A. Ashtekar, ``New variables for classical and quantum gravity,'' {\it{Phys. Rev. Lett.}} vol. 57, 1986, 2244.



\bibitem{aps1} A.~Ashtekar, T.~Pawlowski and P.~Singh,
  ``Quantum nature of the big bang,''
  {\it{Phys.\ Rev.\ Lett.}} vol. 96, 2006, p. 141301.

\bibitem{aps3} A.~Ashtekar, T.~Pawlowski and P.~Singh,
  ``Quantum Nature of the Big Bang: Improved dynamics,''
  {\it{Phys.\ Rev.\ D}} vol. 74, 2006, p. 084003.

\bibitem{as-status}  A.~Ashtekar and P.~Singh,
  ``Loop Quantum Cosmology: A Status Report,''
  {\it{Class.\ Quant.\ Grav.}} vol. 28, 2012, p. 213001.

\bibitem{numlqc1} D.~Brizuela, D.~Cartin, and G.~Khanna, ``Numerical Techniques in Loop Quantum Cosmology,''
 {\it{SIGMA}}, vol. 8, 2012, p. 001.

\bibitem{numlqc2} P.~Singh,
  ``Numerical loop quantum cosmology: an overview,''
  {\it{Class.\ Quant.\ Grav.}} vol. 29, 2012, p. 244002.
\vskip-0.28cm
{\bibitem{cfl} R. Courant, K. O. Friedrichs, H. Lewy, ``On the patrial difference equations of mathematical physics,'' {\it{IBM Journal}}, vol. 11, 1967, p. 215.}

%\bibitem{consistent} \textcolor{black}{D. Craig, P. Singh, ``Consistent probabilities in loop quantum cosmology,'' {\it Class. Quant. Grav.} vol. 30, 2013, p. 205008.}

\bibitem{chimera} P.~Diener, B.~Gupt and P.~Singh,
  ``Chimera: A hybrid approach to numerical loop quantum cosmology,''
  {\it Class.\ Quant.\ Grav.}\  vol. 31, 2014, p. 025013.

\vskip-0.28cm
 {\bibitem{bkl} V. Belinskii, E. Lifshitz, and I. Khalatnikov, ``Oscillatory
approach to the singular point in relativistic cosmology”, {\it{Sov.  Phys.
Usp.}}, vol. 13, 1971, p. 745.}

\bibitem{numlsu-2} P.~Diener, B.~Gupt and P.~Singh,
  ``Numerical simulations of a loop quantum cosmos: robustness of the quantum bounce and the validity of effective dynamics,''
  {\it Class.\ Quant.\ Grav.}  vol. 31, 2014. p. 105015.

\bibitem{numlsu-3}  P.~Diener, B.~Gupt, M.~Megevand and P.~Singh,
  ``Numerical evolution of squeezed and non-Gaussian states in loop quantum cosmology,''
  {\it Class.\ Quant.\ Grav.}\  vol. 31, 2014, p. 165006.


  \bibitem{craig} D. Craig, ``Dynamical eigenfunctions and critical density in loop quantum cosmology,'' {\it Class. Quant. Grav} vol. 30, 2013, p. 035010.



\bibitem{numlsu-4} P.~Diener, A.~Joe, M.~Megevand and P.~Singh,
  ``Numerical simulations of loop quantum Bianchi-I spacetimes,''
  {\it Class.\ Quant.\ Grav.}\  vol. 34, 2017, p. 094004.


\bibitem{madrid}
M.~Martin-Benito, G.~A.~M. Marugan, and T.~Pawlowski, ``Physical evolution in Loop Quantum Cosmology: The example of vacuum Bianchi-I,''
{\it{Phys. Rev. D}}, vol. 80, 2009, p. 084038.


\bibitem{cactus} G.~Allen, T.~Goodale, and E.~Seidel,
\newblock The cactus computational collaboratory: Enabling technologies for
  relativistic astrophysics, and a toolkit for solving pdes by communities in
  science and engineering,
\newblock in {\em 7th Symposium on the Frontiers of Massively Parallel
  Computation-Frontiers 99}, New York, 1999, IEEE.

\bibitem{cactus_web}
{Cactus} {Computational} {Toolkit},
\newblock {\tt http://www.cactuscode.org}.

\bibitem{xsede} J.~Towns {\em et~al.}, ``XSEDE: Accelerating Scientific Discovery,'' Computing in Science \& Engineering, vol.16, 2014, pp. 62-74.

\bibitem{bh-quantum} A.~Yonika, G.~Khanna and P.~Singh,
  ``Von-Neumann Stability and Singularity Resolution in Loop Quantized Schwarzschild Black Hole,'' \textcolor{black}{{\it{Class. Quant. Grav.}}, vol. 35, 2018, p. 045007.}



\end{thebibliography}
\end{document}